\documentclass[nonacm, sigconf]{acmart}

\AtBeginDocument{%
  \providecommand\BibTeX{{%
    \normalfont B\kern-0.5em{\scshape i\kern-0.25em b}\kern-0.8em\TeX}}}
    
\setcopyright{none}
\copyrightyear{2021}
\acmYear{2022}
\acmDOI{xx.xxxx/xxxxxxx.xxxxxxx}
\usepackage{dirtytalk}
\usepackage{pgfplots}
\usepgfplotslibrary{groupplots}
\usetikzlibrary{matrix}
\pgfplotsset{compat=1.17}

\newcommand{\multiplicationLatencyMultPIMOverSerial}{11}
\newcommand{\sortingLatencyAlamOverSerial}{14}

\begin{document}

\title{PartitionPIM: Practical Memristive Partitions for Fast Processing-in-Memory}

\author{Orian Leitersdorf, Ronny Ronen, and Shahar Kvatinsky}

\begin{abstract}
Digital memristive processing-in-memory overcomes the memory wall through a fundamental storage device capable of stateful logic within crossbar arrays.
Dynamically dividing the crossbar arrays by adding memristive partitions further increases parallelism, thereby overcoming an inherent trade-off in memristive processing-in-memory. The algorithmic topology of partitions is highly unique, and was recently exploited to accelerate multiplication ($\multiplicationLatencyMultPIMOverSerial\times$ with 32 partitions) and sorting ($\sortingLatencyAlamOverSerial\times$ with 16 partitions). Yet, the physical implementation of memristive partitions, such as the peripheral decoders and the control message, has never been considered and may lead to vast impracticality. This paper overcomes that challenge with several novel techniques, presenting efficient practical designs of memristive partitions. We begin by formalizing the algorithmic properties of memristive partitions into serial, parallel, and semi-parallel operations. Peripheral overhead is addressed via a novel technique of \emph{half-gates} that enables efficient decoding with negligible overhead. Control overhead is addressed by carefully reducing the operation set of memristive partitions, while resulting in negligible performance impact, by utilizing techniques such as \emph{shared indices} and \emph{pattern generators}. Ultimately, these efficient practical solutions, combined with the vast algorithmic potential, may revolutionize digital memristive processing-in-memory. 
\end{abstract}

\maketitle

\section{Introduction}
\label{sec:introduction}

Logic and storage are united in processing-in-memory (PIM)~\cite{NDP} solutions to overcome the data-transfer bottleneck~\cite{DarkMemory}. Memristive processing-in-memory~\cite{mMPU} is rapidly emerging as an implementation of real processing-in-memory that utilizes the memristor~\cite{Memristor}, a device inherently capable of both storage and logic. Yet, the logic supported by memristive processing-in-memory is basic (a logic gate such as NOR~\cite{MAGIC}), requiring many cycles for arithmetic operations (e.g., 320 cycles for 32-bit addition~\cite{Bitlet}, 11264 cycles for 32-bit multiplication~\cite{Ameer}). This leads to a trade-off between latency and throughput, diminishing the benefit of memristive processing-in-memory. Recent works have overcome this trade-off by proposing the concept of \emph{memristive partitions}~\cite{FELIX} that increases parallelism to achieve arithmetic that is both fast and high-throughput~\cite{MultPIM, RIME, alam2021sorting}. Nevertheless, the implementation of partitions was never discussed, and naive solutions lead to vast impracticality. This paper utilizes several novel techniques to design practical memristive partitions that maintain the immense algorithmic potential, thereby advancing memristive processing-in-memory towards a practical solution that may vastly accelerate large-scale applications. 

\begin{figure}[!t]
\centering 
\includegraphics[width=3in, trim={0cm, 0.4cm, 0cm, 0cm}]{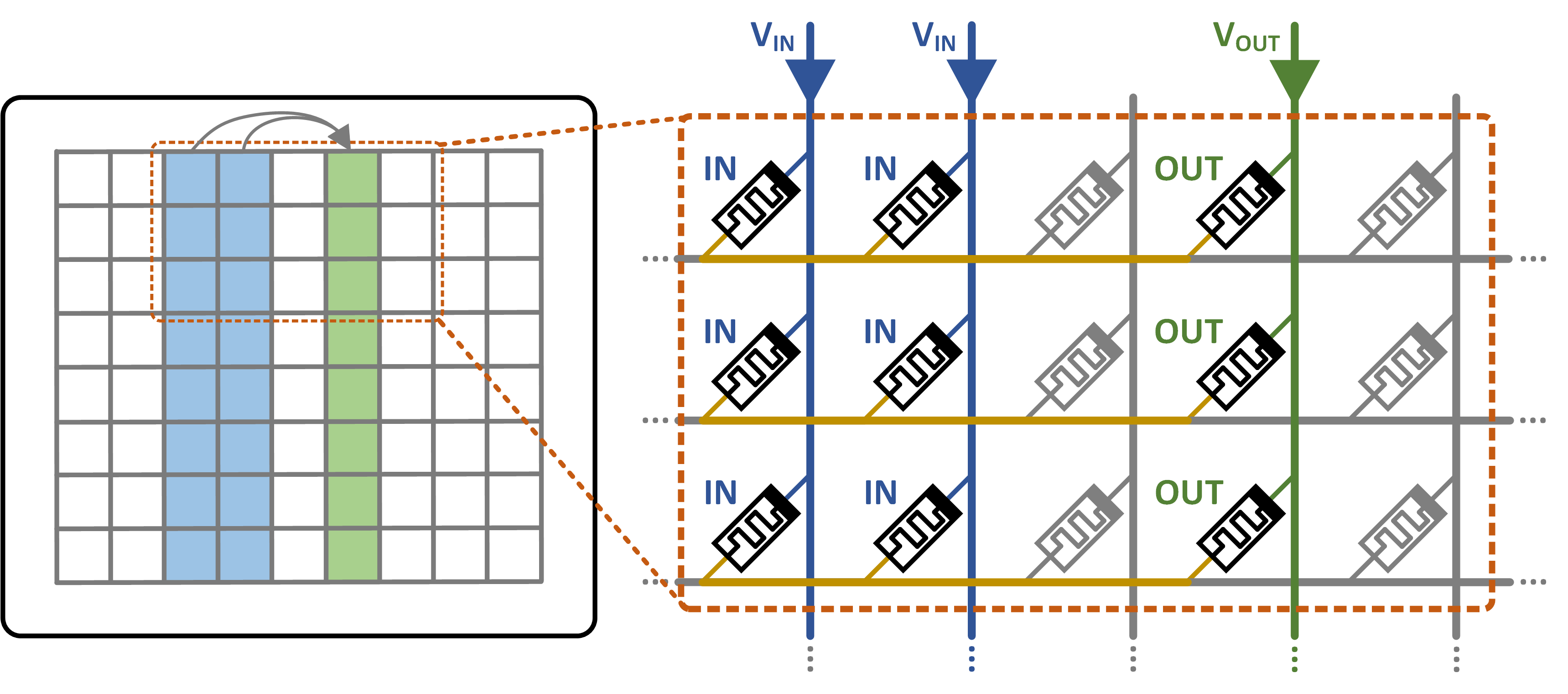}
\caption{Stateful logic within a crossbar with parallelism across all rows by applying voltages across the bitlines.}
\label{fig:crossbar}
\vspace{-15pt}
\end{figure}

The emerging stateful-logic technique enables parallel row or column gates within a memristive crossbar array with constant time. Binary information is represented by the resistance of a memristor, and \emph{stateful logic} enables logic amongst memristors~\cite{MemristiveLogic, borghetti2010memristive, MAGIC, FELIX}. The crossbar array stores a single bit per memristor, and applying voltages across bitlines (wordlines) of the crossbar induces parallel logic within rows (columns). For example, the bit-wise NOR of two columns can be computed and stored in a third column, within a single cycle, by applying only three voltages on the bitlines, as illustrated in Figure~\ref{fig:crossbar}~\cite{MAGIC}. Examples of stateful-logic techniques include MAGIC~\cite{MAGIC} and FELIX~\cite{FELIX}, enabling various logic gates, such as NOT, NOR, NAND, OR, and Minority3, in a single cycle. 

In single-row arithmetic, each row performs computation independently on different input. As such, column operations enable parallel vectored execution across all rows~\cite{SIMPLER}, independent of the vector dimension (row count). Due to its potential high throughput, such arithmetic gained recent attention, including single-row addition~\cite{MultPIM, FELIX, Bitlet, SIMPLER, abstractPIM} and multiplication~\cite{Ameer, FloatPIM, IMAGING}. While logic is parallelized across rows, each gate in a single-row algorithm occurs serially~\cite{MemristiveLogic}. For example, element-wise multiplication of two $n$-dimensional vectors of $N$-bit numbers requires serial execution of $O(N^2)$ gates, hence $O(N^2)$ latency (independent of $n$)~\cite{Ameer}.

\begin{figure*}[!t]
\centering 
\includegraphics[width=6.2in, trim={0cm, 0.2cm, 0cm, 0cm}]{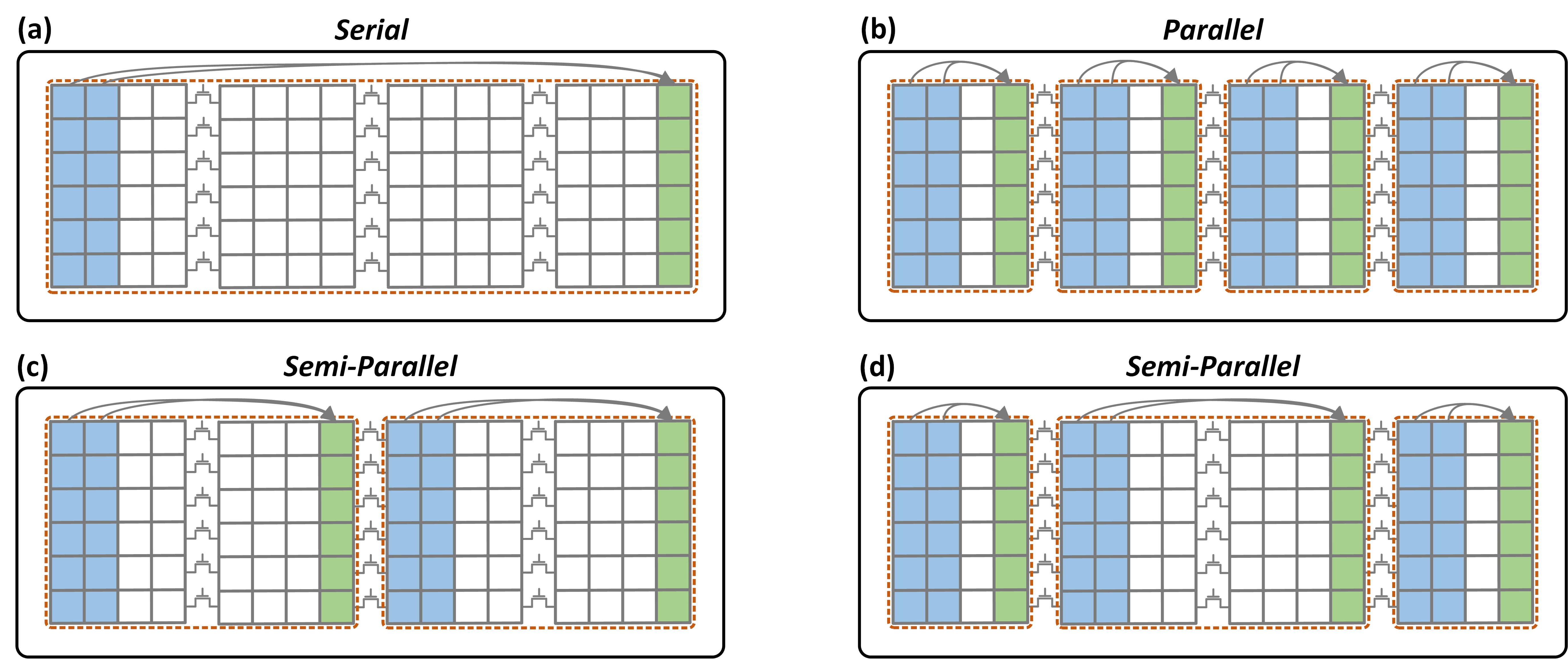}
\caption{Memristive crossbar with row partitions, and the different types of partition-based computation: (a) serial, (b) parallel, and (c,d) semi-parallel. The dynamic section division is shown in dashed orange, inputs are blue, and outputs are green.}
\label{fig:parallelism} 
\vspace{-10pt}
\end{figure*}

Memristive partitions~\cite{FELIX} accelerate stateful logic algorithms by enabling concurrent row or column operations using transistors that divide crossbar wordlines (bitlines) into independent \emph{partitions}. Voltages are applied on the bitlines (wordlines), yet the transistors ensure isolation between the wordlines (bitlines) of distinct stateful-logic gates in the same row (column) to enable concurrent operation. For single-row arithmetic, partitions enable execution of multiple parallel gates \emph{per row}, alleviating the single-gate per cycle constraint, while still executing in parallel across all rows~\cite{FELIX, MultPIM, RIME, alam2021sorting}.

Since the proposition of partitions~\cite{FELIX}, algorithmic works~\cite{MultPIM, RIME, alam2021sorting, FELIX, ICECS} exploited partitions to accelerate single-row multiplication by $\multiplicationLatencyMultPIMOverSerial\times$ (using 32 partitions)~\cite{MultPIM}\footnote{To isolate the effect of partitions, this result compares MultPIM~\cite{MultPIM} to its optimized serial implementation. Note that algorithmic area (memristor footprint) and energy (total gate count) are increased by $1.4\times$ and $2.1\times$, respectively. See Section~\ref{sec:results}.} and sorting by $\sortingLatencyAlamOverSerial\times$ (using 16 partitions)~\cite{alam2021sorting}. MultPIM~\cite{MultPIM} exploited interesting properties that arise when dynamically dividing partitions to develop fast shifting (in constant time) and broadcasting (in logarithmic time) techniques. Preliminary analysis from previous work estimates low area-overhead for the transistors, such as $3\%$ for 32 partitions~\cite{FELIX}. This suggests memristive partitions can, e.g., accelerate multiplication latency by $\multiplicationLatencyMultPIMOverSerial\times$ while only increasing crossbar area by $1.03\times$. Yet, the peripheral circuits and control to support partitions were never previously discussed. The peripheral circuits relate to decoders that apply voltages across bitlines and wordlines, while control refers to controller messages sent to crossbars to convey operations. Without efficient designs of these, realizing the potential of partitions may be over-optimistic, leading to impractical designs. 

This paper proposes efficient designs of partitions. Section~\ref{sec:unlimited} formalizes the algorithmic potential implied by previous works into \emph{serial, parallel} and \emph{semi-parallel} operations, and proposes efficient periphery fully enabling these operations through a novel technique of \emph{half-gates}. Section~\ref{sec:unlimited} also identifies control overhead as an inherent challenge in this \emph{unlimited} model, which is addressed in Section~\ref{sec:standard} (\emph{standard} model) via intra-partition restrictions and in Section~\ref{sec:minimal} (\emph{minimal} model) via inter-partition patterns. Section~\ref{sec:results} presents the trade-off between control overhead and performance, analyzing multiplication~\cite{MultPIM} as a case study. This paper contributes:
\vspace{-10pt}
\begin{enumerate}
    \item \textbf{Partition Models:} Presents well-defined algorithmic models detailing the capabilities of memristive partitions, categorizing operations as \emph{serial}, \emph{parallel} and \emph{semi-parallel}.
    \item \textbf{Periphery:} Proposes efficient crossbar periphery for each model, based on a technique of \emph{half-gates}, incurring even slightly less overhead than a crossbar without partitions. 
    \item \textbf{Control:} Drastically reduces control-message length by restricting operation sets, in the standard and minimal models, with \emph{shared indices} and \emph{pattern generators}, while causing only minimal impact on partition performance.
    \item \textbf{Results:} Case study of parallel multiplication, showing a trade-off between control overhead and performance. Control overhead is reduced by $17\times$ while latency is only increased by $1.3\times$ (that is, $9\times$ latency over a baseline with no partitions rather than the theoretical $11\times$). 
\end{enumerate}

\section{Unlimited Design}
\label{sec:unlimited}

\begin{figure*}[!th]
\centering 
\includegraphics[width=0.94\linewidth]{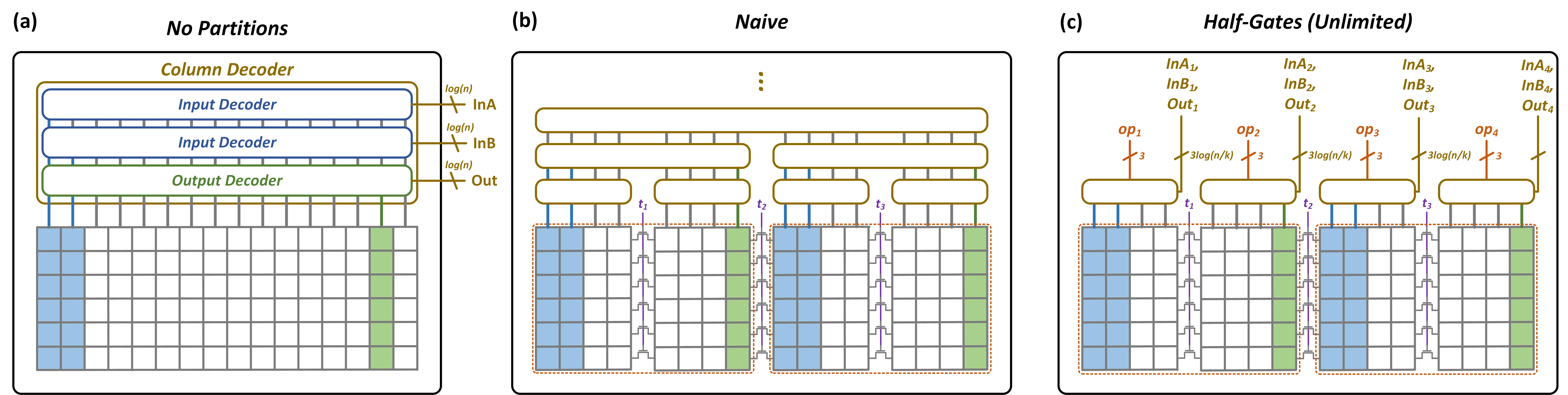}
\caption{(a) The periphery for a baseline crossbar with no partitions, with each decoder unit consisting of analog multiplexers (per bitline, not shown in the figure) and CMOS decoders~\cite{NishilThesis}. (b) A naive impractical approach at utilizing the column decoder for the unlimited model, and (c) the proposed approach based on the \emph{half-gates} technique.}.
\label{fig:unlimited} 
\vspace{-10pt}
\end{figure*}

We formalize the operations enabled by partitions into serial, parallel, or semi-parallel; the unlimited model supports all of the possibilities for each. Semi-parallel operations recently emerged~\cite{MultPIM}, enabling efficient communication between partitions that significantly improves results over only serial and parallel partition parallelism, such as a $4\times$ latency improvement in multiplication~\cite{MultPIM, RIME}. We then detail a naive approach to crossbar periphery for the unlimited model, which we replace with an efficient solution based on the novel \emph{half-gates} technique. Finally, we identify control overhead as a critical inherent challenge in the unlimited model, which is resolved in Sections~\ref{sec:standard} and \ref{sec:minimal} by carefully reducing the operation set.

\vspace{-5pt}

\subsection{Model}
\label{sec:unlimited:model}

Memristive partitions enable a unique parallelism that may be exploited for efficient techniques. Consider inserting $k-1$ transistors at fixed locations into each row of the $n \times n$ crossbar, as illustrated in Figure~\ref{fig:parallelism}. The transistors dynamically isolate different parts of each row to enable concurrent execution, essentially dynamically dividing the crossbar partitions into \emph{sections} (dashed orange) such that each section may perform a column operation. Initial works~\cite{FELIX, RIME, alam2021sorting} utilized partitions in a binary fashion: either each partition is a section (parallel), or the entire crossbar is one section (serial). A recent work~\cite{MultPIM} demonstrated the potential of semi-parallelism, significantly improving solutions that utilize only serial and parallel, e.g., $4\times$ improvement in latency for multiplication~\cite{MultPIM, RIME}. We define these various parallelism forms:
\begin{itemize}
    \item \textbf{Serial} (Figure~\ref{fig:parallelism}(a)): When the transistors are \emph{all conducting}, then the crossbar is equivalent to one without partitions. Therefore, only a single gate is operated per cycle. 
    \item \textbf{Parallel} (Figure~\ref{fig:parallelism}(b)): When the transistors are \emph{all not conducting}, then $k$ gates may operate concurrently as part of an operation, \emph{one gate within each partition}. 
    \item \textbf{Semi-Parallel} (Figures~\ref{fig:parallelism}(c,d)): When only \emph{some} transistors are \emph{conducting}, then multiple gates may operate concurrently, \emph{between partitions}. Essentially, an operation is a set of \emph{disjoint} intervals that represent the underlying partitions of sections. 
\end{itemize}

\subsection{Periphery}
\label{sec:unlimited:periphery}

We propose a low-overhead peripheral design that supports the unlimited model. We begin with a short background on the peripheral circuits in a crossbar without partitions, and then discuss a naive approach for partition periphery that is not practical. Conversely, we introduce a unique technique of \emph{half-gates}, which enables practical partition periphery with very low overhead.

Traditional crossbar periphery applies voltages across bitlines or wordlines to induce vectored logic. Without loss of generality, we discuss only bitlines and assume a single two-input gate type (e.g., NOR)\footnote{The proposed designs can be generalized to support additional types of gates (e.g., NAND, OR), including gates with more than two inputs~\cite{FELIX}.}. The periphery for such a crossbar consists of a column decoder that receives indices $InA$, $InB$, and $Out$, and applies $V_{IN}$ on bitlines $InA$/$InB$ and $V_{OUT}$ on bitline $Out$, as illustrated in Figure~\ref{fig:unlimited}(a). The column decoder is composed of three decoder units, each of which receives a single index and outputs a fixed voltage at that index. Each decoder unit consists of a CMOS decoder and an analog multiplexer for each bitline: the CMOS decoder provides the selection to the analog multiplexers of each bitline~\cite{NishilThesis, MemristiveLogic, mMPUController}. 

The naive approach to supporting the unlimited model utilizes the overall column decoder structure for enabling serial, parallel, and semi-parallel operations. For supporting serial operations, a single column decoder across all bitlines is necessary. For supporting parallel operations, a column decoder is needed for every partition individually. For semi-parallel operations, decoders are needed for any possible configuration of the sections. Overall, the naive approach requires a \emph{stack} of $\Omega(k^2)$ column decoders (for the unlimited model), which is clearly not practical, see Figure~\ref{fig:unlimited}(b). Note that this is in addition to signals that control the transistors. Other naive solutions will likely also suffer from this overhead.

\begin{table}[!t]
\caption{Opcode for an Individual Partition}
\centering
\begin{tabular}{c|c|c|c}
 Index & Opcode & Index & Opcode \\
 \hline
 000 & - & 100 & $Gate(InA, ?) \rightarrow \; ?$ \\
 001 & $? \rightarrow Out$ & 101 & $Gate(InA, ?) \rightarrow Out$ \\
 010 & $Gate(?, InB) \rightarrow \; ?$ & 110 & $Gate(InA, InB) \rightarrow \; ?$ \\
 011 & $Gate(?, InB) \rightarrow Out$ & 111 & $Gate(InA, InB) \rightarrow \; Out$ \\
\end{tabular}
\label{table:opcodes}
\vspace{-10pt}
\end{table}

\begin{figure}[!t]
\centering 
\includegraphics[width=3.2in]{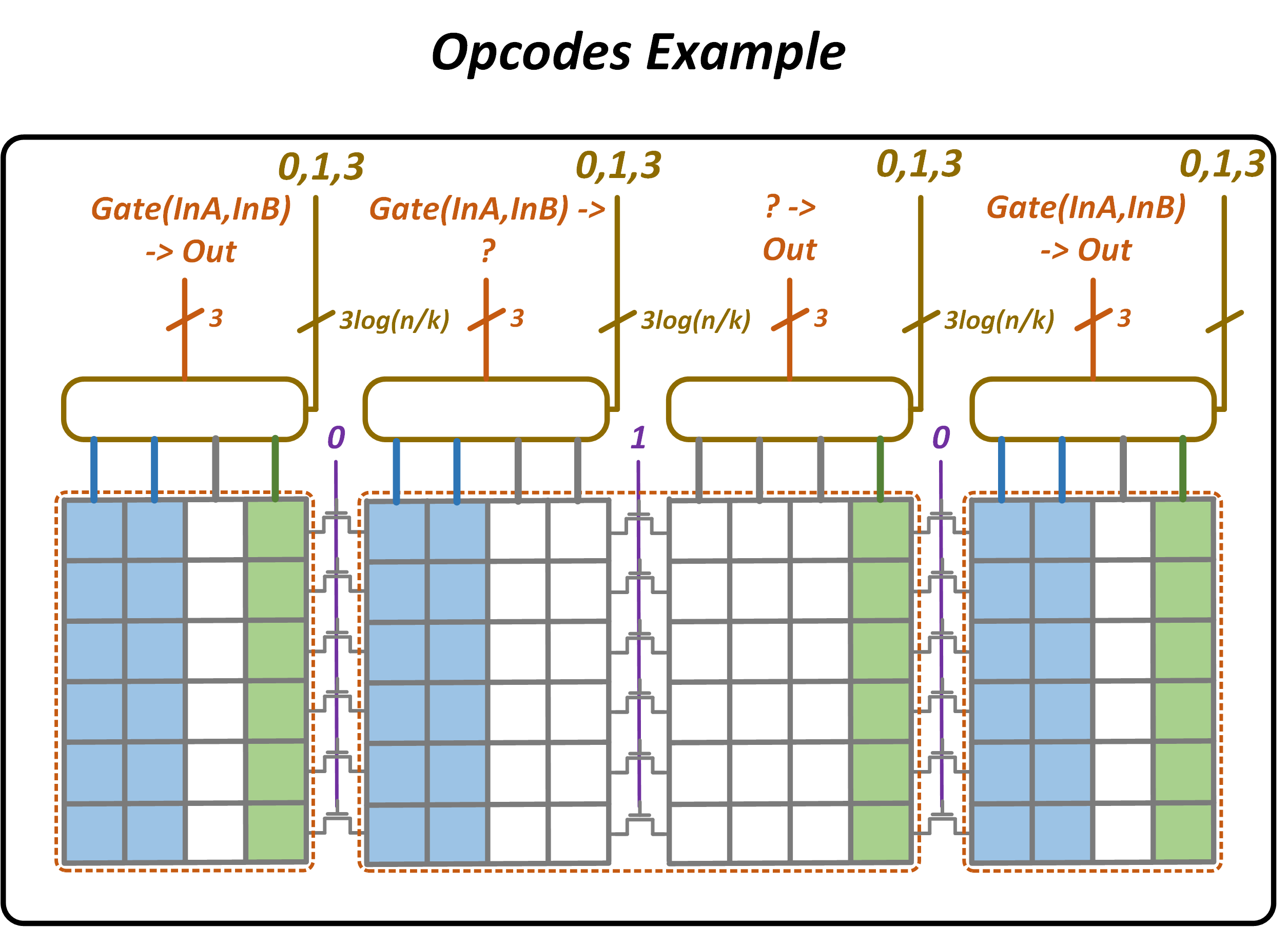}
\caption{Demonstration of the opcodes for each partition in the half-gate technique for the operation from Figure~\ref{fig:parallelism}(d).}
\label{fig:opcodes} 
\vspace{-10pt}
\end{figure}

Our proposed approach is based on \emph{half-gates}: we eliminate the stack of column decoders, utilize only a single column decoder per partition, yet introduce an opcode for each partition, as shown in Figure~\ref{fig:unlimited}(c). We describe the basic idea through the following example: to support a gate where inputs are in partition $p_1$ and outputs are in partition $p_2$ (both in the same section), (1) the column decoder of $p_1$ applies only the input voltages without applying the output voltages, and (2) the column decoder of $p_2$ applies only the output voltages without applying the input voltages. Essentially, $p_1$ trusts that a different partition will apply output voltages, and $p_2$ trusts that a different partition will apply input voltages. While each gate on its own is not valid (\emph{half-gate}), their combination is valid. Table~\ref{table:opcodes} details the various possible opcode states of each column decoder, where \say{?} represents not applying voltages for that part of the gate and \say{-} represents not applying voltages at all (for intermediate partitions). An example opcode setting is shown in Figure~\ref{fig:opcodes} for the operation from Figure~\ref{fig:parallelism}(d).  Note that the opcode decoding is simple: two bits are the enables for the input decoder units, and the other bit is the enable for the output decoder unit.

In terms of the decoder structure, the proposed design is nearly identical to the baseline crossbar. In essence, the proposed decoder is identical to concatenating multiple baseline $n/k$-column decoders horizontally (thus width remains at $O(k \cdot (n/k)) = O(n))$). The analog multiplexers within the decoders remain completely identical, and the only difference is the CMOS decoders that generate the selects for the analog multiplexers. Rather than a CMOS $n$-multiplexer, the proposed solution requires $k$ CMOS $n/k$-multiplexers. In terms of CMOS gates, the proposed solution requires \emph{less} gates than the baseline crossbar as $\log_2(n/k) < \log_2(n)$. This saving in gates comes at the cost of increased control complexity. 

\subsection{Control}
\label{sec:unlimited:control}

The proposed periphery decodes a relatively-long message sent from the controller that details the operation. 
We demonstrate that this length is nearly optimal for the unlimited model. The proposed peripheral decoding requires $3k \cdot \log_2(n/k) + 3k + (k-1)$ bits to encode an operation that may occur in a single cycle (indices, opcodes, and transistor selects, respectively), for $k$ evenly-spaced partitions amongst $n$ bitlines. For $k=32$ and $n=1024$, this requires $607$ bits, compared to the $30$ bits required in a crossbar without partitions. This $20\times$ difference is concerning as the communication architecture between the controller and the crossbars must support $20\times$ larger messages, incurring massive area and energy overhead. We prove that this message length is nearly optimal (and not as a result of poor decoding) through a combinatorial analysis:
\begin{itemize}
    \item \emph{Serial:} There are $\binom{n}{2} \cdot (n-2)$ different possible serial operations (choices for $InA, InB$ and $Out$).
    \item \emph{Parallel:} There are $\big [\binom{n/k}{2} \cdot (n/k-2)\big]^k$ possible parallel operations (each partition performs a different gate).
    \item \emph{Semi-Parallel:} 
    We do not count semi-parallel operations for unlimited. This is valid as we seek a lower-bound.
\end{itemize}

We find over $2^{443}$ different operations, thus $\log_2\left(2^{443}\right)$ is a lower-bound on the message length. Therefore, any implementation of the unlimited model will require a length of at least $443$ bits (compared to $30$ without partitions). Note that $443$ is not very far from $607$, and that attaining this lower-bound likely requires complex decoding. This challenge was not considered by the previous works~\cite{FELIX}, and is addressed in the next sections. 

\vspace{-5pt}

\section{Standard Design}
\label{sec:standard}

The standard model addresses the control-overhead challenge with the unlimited model. We seek to significantly reduce the control-message length, while having low impact on performance. We still support serial, parallel, and semi-parallel operations, yet we eliminate the less-common operations primarily by restricting allowed \emph{intra-partition} patterns.

\subsection{Model}
\label{sec:standard:model}

We set the following criteria in addition to Section~\ref{sec:unlimited:model}:

\begin{itemize}
    \item \emph{Identical Indices:} The input and output \emph{intra-partition} indices of the gates must be identical. Figure~\ref{fig:parallelism}(d) is supported as the indices within each partition are identical (in the example, inputs are always first/second columns, output is always the fourth column). That is, for evenly distributed partitions, the indices modulo $n/k$ must be identical.
    \item \emph{No Split-Input:} Gate inputs $InA$ and $InB$ must belong to the same partition, for each gate\footnote{This affects serial operations as well. Serial algorithms may overcome this limitation by copying one of the inputs, or by adjusting the mapping algorithms~\cite{SIMPLER}.}.
    \item \emph{Uniform Direction:} The direction (\say{inputs left of outputs}, or \say{outputs left of inputs}) is identical in all concurrent gates for semi-parallel operations.
\end{itemize}

Note that all of the examples in Figure~\ref{fig:parallelism} are supported, as well as the general-purpose partition techniques from MultPIM~\cite{MultPIM}\footnote{While the techniques are supported, the original MultPIM algorithm slightly violates the \emph{Identical Indices} criteria with the first/last partitions. Section~\ref{sec:results} replaces those illegal operations with supported alternatives as part of the evaluation.}.

\vspace{-5pt}

\subsection{Periphery}
\label{sec:standard:periphery}

The peripheral modifications from the unlimited model to the standard model are two-fold: the intra-partition indices are shared (following the \emph{Identical Indices} criteria), and the opcodes are automatically generated from the section division (following the \emph{No Split-Input} and \emph{Uniform Direction} criteria). The periphery is slightly reduced compared to that of a baseline crossbar without partitions, suggesting low overhead.

\vspace{5pt}

\subsubsection{Intra-Partition Indices}
The modification from unlimited is rather simple: the indices provided to each decoder are identical (see Figure~\ref{fig:standard}). Note that this enables an additional optimization: shared CMOS decoders. Recall that each of the column decoders is composed of input/output decoder units, and that each such unit is composed of a CMOS decoder and analog multiplexers (for each bitline) -- the CMOS decoder provides the select lines for the analog multiplexers~\cite{NishilThesis}. Given that the indices are now shared across partitions, the CMOS decoders may be shared while the analog multiplexers remain unchanged (compared to a baseline crossbar). Therefore, this has the potential to further \emph{reduce} the proposed peripheral overhead compared to a baseline crossbar.

\vspace{5pt}

\subsubsection{Opcode Generation} The \emph{No Split-Input} and \emph{Uniform Direction} criteria, combined with a unique observation, enable opcode derivation from the section division and a single enable bit per partition. First, we note that there may exist multiple valid section divisions for a single semi-parallel operation; for example, any semi-parallel operation that does not utilize the first partition may set the first transistor to either conducting or non-conducting. We restrict this degree of freedom by defining a \emph{tight} section division as one satisfying that no section can be split; e.g., the first transistor would be non-conducting in the previous example. In a tight section division, the first and last partitions of each section always contain either an input or output, and the middle partitions (if exist) are always unused. The only exception is sections that do not contain any gate. Therefore, given only the section division, the direction of the operation, and an indication of whether each section contains a gate, the opcodes of all partitions may be generated. 

\begin{figure}[!t]
\centering 
\includegraphics[width=2.9in]{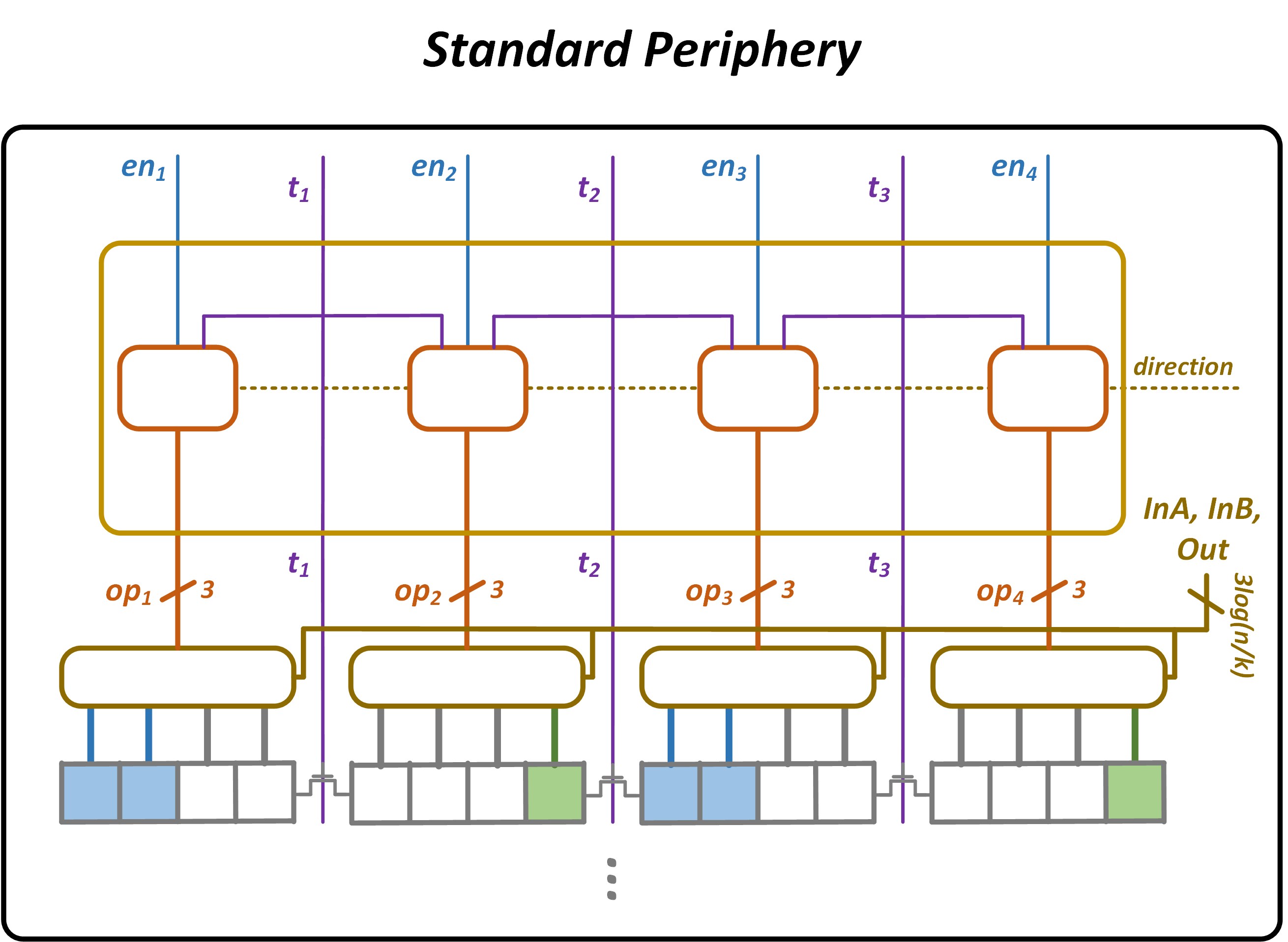}
\caption{The periphery of the standard model, sharing the indices across the decoders and generating opcodes (each orange box consists of two 2:1 multiplexers, see Section~\ref{sec:standard:periphery}).
}
\vspace{-15pt}
\label{fig:standard} 
\end{figure}

Opcode generation is achieved following this observation, as illustrated in Figure~\ref{fig:standard}. Given the transistor selects (which are chosen to define a \emph{tight} section division), an enable bit for each partition, and a general direction bit (\say{inputs left of outputs}, or \say{outputs left of inputs}), the opcode generator computes the opcodes of all partitions (which are inputted to the decoders). For the direction of \say{inputs left of outputs}, the input bits of an opcode are logical one if the transistor to the left of that partition is selected, and the output bits are logical one if the transistor to the right is selected (vice-versa for \say{outputs left of inputs}) -- with the opcode ANDed with the enable of that partition. Therefore, the opcode for a partition may be derived from the select of the transistors to the left and right, the enable of that partition, and the general direction. Such decoding may be achieved by two 2:1 multiplexers per partition (only $O(k)$ gates total, negligible compared to $O(n\log k)$ gates elsewhere). 

\subsection{Control}
\label{sec:standard:control}
The message-length of the unlimited model is far reduced in the standard model by index sharing and opcode generation. Standard decoding uses $3 \cdot \log_2(n/k) + (2k-1) + 1$ bits (indices, enables and transistor selects, and direction, respectively), nearly eliminating the bottleneck of unlimited ($3k \cdot \log_2(n/k)$), and mildly reducing the rest ($4 k-1$). Message length is reduced from $607$ to $79$ bits -- a $7.7\times$ improvement (for $k=32$ and $n=1024$). From a combinatorial analysis, we find at least $2 \cdot \sum_{m=1}^k \binom{k-1}{m-1} \binom{n/k}{2}\cdot (n/k-2)$ supported operations, thus a $46$ bit lower-bound -- not very far from $79$ bits.

\section{Minimal Design}
\label{sec:minimal}
The control overhead in Section~\ref{sec:standard} remains relatively high due to the transistor selects and enables: any section division is supported. The minimal design addresses that concern by requiring that the section division follows several carefully-chosen \emph{inter-partition} patterns.

\subsection{Model}
\label{sec:minimal:model}
The standard model supports any division of partitions into sections, but in practice, many divisions are typically not used. For example, Figure~\ref{fig:parallelism}(d) is rarely used -- e.g., not at all in MultPIM. We identify two criteria typically followed:
\begin{itemize}
    \item \emph{Uniform Partition-Distance:} The \emph{partition distance} of a gate is the distance between its input and output partitions (e.g., (1,1) for Figure~\ref{fig:parallelism}(c); (0,1,0) for Figure~\ref{fig:parallelism}(d)). Gates performed concurrently should all have identical partition distance.
    \item \emph{Periodic:} The gates must repeat periodically, e.g., every $T$ partitions (for $T$ greater than the partition distance). 
\end{itemize}
Examples (a), (b), and (c) from Figure~\ref{fig:parallelism}  are supported, as well as the partition techniques from MultPIM~\cite{MultPIM}\footnote{While the techniques are supported, the MultPIM algorithm slightly violates the \emph{Periodic} criteria, see Section~\ref{sec:results} for the alternative.}. Note that typical usage of partitions already follows the above restrictions, suggesting the minimal model is general-purpose.

\subsection{Periphery}
\label{sec:minimal:periphery}
The decoder for the minimal design replaces the opcode generator from standard, following these key observations:
\begin{itemize}
    \item Input opcodes can be derived from a \emph{Range Generator}, outputting logical one every period $T$, from $p_{start}$ to $p_{end}$. This may be accomplished with two shifters (for $p_{start}$ and $p_{end}$) and a decoder (for $T$). 
    \item Output opcodes can be derived by shifting the input opcodes by the partition distance according to the global direction (up to $k$ shift in either direction).
    \item Transistor selects can be derived from input and output opcodes. For example, if the global direction is \say{input left of output}, then a separation transistor is non-conducting if there is output to its left or input to its right.
\end{itemize}

The overall periphery is similar to that of standard, while replacing the opcode generator with the above shifters and decoder. 
Note the periphery overhead here is relatively low as shifters and decoder operate on width $k$ (rather than $n$).

\subsection{Control}
\label{sec:minimal:control}
The moderate-length control message in the standard model is drastically reduced in the minimal model, attaining $3 \cdot \log_2(n/k) + 3 \cdot \log_2(k) + \log_2(k) + 1$ bits (intra-partition indices, range indices, partition-distance, and global direction, respectively). For $n=1024$ and $k=32$, this improves from 607 bits (unlimited) and 79 bits (standard) to only 36 bits. Interestingly, this small message is still capable of supporting most of the operations used in algorithms, see Section~\ref{sec:results}. We find a lower bound of at least 25 bits from all non-input-split serial operations being supported.

\section{Evaluation}
\label{sec:results}

We analyze the unlimited, standard, and minimal models, presenting a trade-off between overhead and performance. We evaluate the effect of standard and minimal on performance by examining their effect on MultPIM~\cite{MultPIM}, as a case study.
While the proposed mechanisms can be generalized to three-input gates (e.g., Minority3), we assume up to two inputs for simplicity and thus consider the NOT/NOR implementation of MultPIM. While the partition techniques proposed in MultPIM are supported by standard and minimal models, MultPIM includes various operations that are not supported. Those operations are replaced with alternatives that are compatible, yet require additional latency -- the details are provided as part of the modified cycle-accurate simulations\footnote{Available at  \url{https://github.com/oleitersdorf/PartitionPIM}.}.

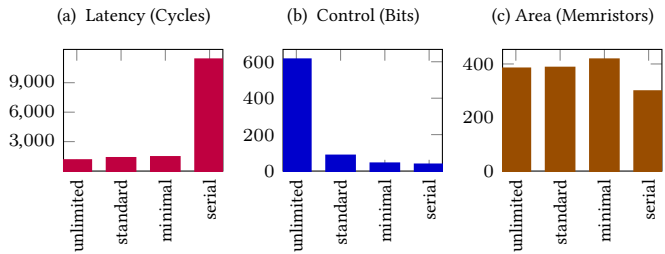
\begin{figure} 
    \centering
  \begin{tikzpicture}
    \begin{groupplot}[
      group style={group size=3 by 1, horizontal sep=0.8cm},
      width=3.7cm, height=3.2cm
    ]
    \nextgroupplot[
        scaled y ticks=false,
        symbolic x coords={unlimited,standard,minimal, serial},
        x tick label style={rotate=90},
        title={\footnotesize (a) \ Latency (Cycles) \ \ \ \ },
        ytick={3000, 6000, 9000},
        ymin=0,
        every axis plot/.append style={ultra thick},
        every tick label/.append style={font=\footnotesize}
    ]
    \addplot[ybar,color=purple, fill=purple] coordinates {
        (unlimited,995) (standard,1219) (minimal,1316) (serial,11264)
    }; 
    
    \nextgroupplot[
        symbolic x coords={unlimited,standard,minimal, serial},
        x tick label style={rotate=90},
        title={\footnotesize (b)\ \ Control (Bits) \ \ \ \ },
        ymin=0,
        every axis plot/.append style={ultra thick},
        every tick label/.append style={font=\footnotesize}
    ]
    \addplot[ybar,color=blue!80!black, fill=blue!80!black] coordinates {
        (unlimited,607) (standard,79) (minimal,36) (serial,30)
    }; 
    
    \nextgroupplot[
        symbolic x coords={unlimited,standard,minimal, serial},
        x tick label style={rotate=90},
        title={\footnotesize (c) Area (Memristors) \ \ \ \ },
        ymin=0,
        every axis plot/.append style={ultra thick},
        every tick label/.append style={font=\footnotesize}
    ]
    \addplot[ybar,color=orange!60!black, fill=orange!60!black] coordinates {
        (unlimited,379) (standard,382) (minimal,413) (serial,294)
    }; 
    
    \end{groupplot}
  \end{tikzpicture}
  \caption{Comparison of (a) latency, (b) control overhead, and (c) algorithmic area for 32-bit multiplication under the proposed partition models and a serial baseline.}
  \label{fig:results} 
  \vspace{-10pt}
\end{figure}

We demonstrate $9\times$ latency improvement for multiplication with the proposed minimal model compared to an optimized serial algorithm, requiring only approximately $1.4\times$ area and $1.2\times$ control overhead. Figure~\ref{fig:results} shows the results for 32-bit multiplication, comparing latency, control-overhead, and area.

\subsection{Latency}
The standard and minimal implementations of MultPIM incur, relative to unlimited, a slight latency increase of $1.23\times$ and $1.32\times$, respectively, as seen in Figure~\ref{fig:results}(a). Yet, this latency is still $9.2\times$ and $8.6\times$ faster compared to an optimized serial multiplier, respectively. 

\subsection{Control}

The control overhead results follow from the expressions derived in the previous sections, see Figure~\ref{fig:results}(b). Assuming $k=32$ and $n=1024$ with gate-types of NOT/NOR, the control-message length for unlimited is 607 bits, 79 bits for standard, and 36 bits for minimal -- compared to 30 bits for a baseline crossbar without partitions. That is, through the novel techniques (shared indices and pattern generators) and restrictions, the minimal model achieves a low control overhead of only $1.2\times$, compared to $20\times$ for unlimited.

\subsection{Area}
Area overhead is composed of physical overhead (mentioned in the previous sections), and algorithmic overhead that arises from requiring additional intermediate memristors within the crossbar.

\subsubsection{Physical Overhead}
The half-gates technique, utilized in all three models, achieves low peripheral overhead compared to a baseline crossbar (without partitions). In fact, the peripheral complexity of the proposed solutions is slightly \emph{lower} than that of a baseline crossbar as the decoder widths are smaller, see Section~\ref{sec:unlimited:periphery}. Furthermore, the crossbar and the analog multiplexers remain completely unchanged. While additional logic is required in the standard and minimal solutions for decoders and shifters, that overhead is relatively low considering that they operate on width $k$ and not $n$. Exact results depend highly on the exact crossbar structure (e.g., 1R, 1S1R, 1T1R), requiring a full physical design of a crossbar array and periphery. Regardless, considering the fact that the peripheral complexity is slightly \emph{decreased} compared to the baseline, we conclude that peripheral area overhead is negligible compared to algorithmic area overhead (and the potential $20\times$ control overhead).

\subsubsection{Algorithmic Overhead}
Algorithmic area overhead, shown in Figure~\ref{fig:results}(c), is based on the number of memristors required. All parallel approaches have higher area overhead than the serial approach as utilizing parallel operations requires intermediates per partition rather than per crossbar. The minor differences in the area overhead between the unlimited, standard and minimal approaches originate from the alternatives to the unsupported operations.

\subsection{Energy}
Energy consumption for stateful-logic is dominated by the memristor switching energy~\cite{NishilThesis}. Therefore, energy is approximated by the total gate count~\cite{Bitlet}. For MultPIM, the energy overhead is $2.1\times$ from serial to parallel: while the latency is improved, more gates occur due to the partition parallelism.

\section{Conclusion}
\label{sec:conclusion}
The algorithmic potential of emerging memristive partitions is highly unique, and may advance digital memristive processing-in-memory by overcoming an inherent trade-off that leads to slow arithmetic operations. Nevertheless, the physical design of partitions has never been discussed, and naive designs for periphery and control may incur massive overhead that leads to vast impracticality. This paper proposes efficient periphery and control through three potential designs with varying flexibility: unlimited, standard, and minimal. We demonstrate efficient periphery for all three models by utilizing a novel technique of \emph{half-gates}, yet identify control overhead as an inherent concern in the unlimited model. We drastically reduce this control overhead by carefully minimizing the operation set while resulting in negligible performance impact, utilizing techniques such as \emph{shared indices} and \emph{pattern generators}. Through a case study of multiplication, we conclude that the proposed practical designs of partitions, coupled with the previous algorithmic works, suggest that partitions will be a crucial element in the integration of memristive processing-in-memory in computing devices.

\section*{Acknowledgment}
This work was supported in part by the European Research Council through the European Union's Horizon 2020 Research and Innovation Programe under Grant 757259, and in part by the Israel Science Foundation under Grant 1514/17.

\bibliographystyle{ACM-Reference-Format}
\bibliography{refs}

\end{document}